\documentclass[jkps,twocolumn,fleqn,showpacs,showkeys]{revtex4}
\usepackage{graphicx}
\usepackage{amssymb}
\usepackage{amsmath}
\usepackage{amsfonts}
\usepackage[T1]{fontenc}
\usepackage{tgbonum}
\usepackage{siunitx}
\usepackage{newtxtext,newtxmath}
\usepackage{xcolor}
\usepackage{bm}

\begin{document}
\title[]{Quantifying Traffic Patterns with Percolation Theory: A Case Study of Seoul Roads}
\author{Yongsung \surname{Kwon}}

\affiliation{Department of Applied Artificial Intelligence, Hanyang University, Ansan 15588, Korea}
\author{Mi Jin \surname{Lee}}
\email{mijinlee@hanyang.ac.kr}
\affiliation{Department of Applied Physics, Hanyang University, Ansan 15588, Korea}
\author{Seung-Woo \surname{Son}}
\email{sonswoo@hanyang.ac.kr}
\affiliation{Department of Applied Physics, Hanyang University, Ansan 15588, Korea}
\affiliation{Department of Applied Artificial Intelligence, Hanyang University, Ansan 15588, Korea}

\begin{abstract}
Urban traffic systems are characterized by dynamic interactions between congestion and free-flow states, influenced by human activity and road topology. This study employs percolation theory to analyze traffic dynamics in Seoul, focusing on the transition point $q_c$ and Fisher exponent $\tau$. The transition point $q_c$ quantifies the robustness of the free-flow clusters, while the exponent $\tau$ captures the spatial fragmentation of the traffic networks. Our analysis reveals temporal variations in these metrics, with lower $q_c$ and lower $\tau$ values generally during rush hours representing low-dimensional behavior, within the broader context of the positive correlation between $q_c$ and $\tau$. Weight-weight correlations are found to significantly impact cluster formation, driving the early onset of dominant traffic states. Comparisons with uncorrelated models highlight the role of real-world correlations. This approach provides a comprehensive framework for evaluating traffic resilience and informs strategies to optimize urban transportation systems.
\end{abstract}



\maketitle

\section{INTRODUCTION}
In the era of big data, the digitization and database storage of information across diverse fields have made vast amounts of data easily accessible. The immense amount and diversity of collected data pose significant challenges in immediate processing and meaningful utilization. Extracting relevant insights from these data requires suitable methods and approaches to identify and interpret desired trends. This challenge is also evident in traffic systems, which generate a wealth of information reflecting the complexities of modern transportation. Transportation systems, encompassing public and private vehicles, reflect these complexities, with public transit playing a pivotal role in daily life. Analyzing these intricate traffic patterns can uncover insights that improve transportation efficiency and enhance urban living standards~\cite{Norgate02042020,LUNKE2020100842}.

Various approaches have been proposed to understand road traffic, including fluid dynamics~\cite{Suryadjaja_Hutagalung_Sutarto_2020, KANG2022149, doi:10.1142/S0218202502002343}, reaction-diffusion dynamics~\cite{RN90}, and contagion-based processes~\cite{RN91} at macroscopic scales, alongside microscopic models such as cellular automata~\cite{LEE20114555, PhysRevE.108.014302}. Researchers have also identified structural patterns in traffic networks, such as tree-like and loop structures during rush hours~\cite{PhysRevE.108.054312} and bottleneck roads contributing to congestion propagation~\cite{RN92, akbarzadeh2018look,10.1063/5.0150217}. These analyses reveal that traffic systems are influenced by more than just road and vehicle characteristics; factors such as road connectivity, geographic constraints, urban planning, and human activity also play critical roles~\cite{kirkley2018betweenness,LEE2023113770,nagel1995emergent}.

As another approach rooted in statistical physics, percolation theory on complex networks~\cite{sykes1964exact, newman2000efficient, moloney2005complexity, bak2013nature} has been popularly utilized to study the propagation of traffic congestion or breakdown of smooth traffic flows~\cite{RN94,wang2015percolation,zeng2019switch,li2015percolation,HAMEDMOGHADAM2022103922}. Recent studies have explored critical phenomena in traffic systems, such as the transition point and Fisher exponent, to quantify the robustness of free-flow clusters~\cite{zeng2019switch}. By removing roads (links) below a threshold $q$, researchers have observed the fragmentation of free-flow clusters as the threshold $q$ increases. The threshold $q_c$, where the giant connected component (GCC) significantly diminishes, indicates the endurance of smooth traffic conditions. A high $q_c$ signals robust traffic flow, while the Fisher exponent $\tau$ quantifies the distribution of smaller clusters, reflecting their scatteredness across the network. Combining with ordinary percolation theory, they have interpreted that the large $\tau$ means that the road networks behave like the mean field caused by the well-performing shortcuts. Hence, $q_c$ and $\tau$ offer significant insight into traffic resilience and network functionality.
The researchers have applied this analysis to three metropolitan cities in China and found convincing results depending on rush hours or non-rush hours~\cite{zeng2019switch}.

Building on these findings, this study applies percolation analysis to Seoul's road network as a case study, a city known for its complex traffic dynamics and dense urban structure. Seoul has its distinctive road topology, characterized by a north-to-south separation influenced by the Han River. We analyze traffic patterns at both hourly (e.g., rush vs. non-rush hours) and daily (weekdays vs. weekends) scales, providing a comprehensive understanding of temporal variations in traffic resilience. 

This study advances the previous study~\cite{zeng2019switch} by emphasizing the impact of weight-weight correlations. We discover that the transition point (the extent of the endurance of the free-flow clusters) does not alter due to the traffic correlation, but it lowers the value of the Fisher exponent $\tau$ (the distribution of the small free-flow clusters). From this finding, we assert that this newly obtained $\tau$ from uncorrelated cases is deemed as the proper baseline for understanding the traffic situation rather than $\tau$ in two dimensions previously suggested. 

The remainder of this paper is structured as follows: Section~\ref{sec:data} describes the road network and taxi dataset in Seoul. Section~\ref{sec:percolation} contains the percolation method and the percolation analysis, including the transition point and the Fisher exponent. Finally, we close this study by summarizing our findings and discussing the implications in Sec.~\ref{sec:conclusion}.

\section{Empirical data for traffic with percolation analysis}
\label{sec:data}
\subsection{Road network and taxi data of Seoul}
\label{subsec:road_taxi}

Seoul is the capital city of South Korea with a population of 10 million people (20 percent of the total population in South Korea), albeit its area accounts for only 0.6 percent of the nation's total land area. Furthermore, the Han River flows through the city center, dividing Seoul into northern and southern parts with its impressive width of approximately 1\si{km}. These northern and southern parts of Seoul are joined by about thirty bridges. Due to the high population density and the structural feature, this city is rich in complex traffic behaviors. We use Seoul's road network as the embedded structure. The Python package {\fontfamily{cmtt}\selectfont OpenStreetMap networkx (OSMnx)}~\cite{osmnx} provides Seoul's road network data, originally containing 69,565 nodes (intersections) and 198,905 links (roads).

For traffic analysis, we utilize speed data from approximately 20,000 corporate taxis. Unlike other forms of public transportation restricted to designated routes, taxis can traverse the entire city, including narrow alleyways. The data obtained from Seoul Transport Operation and Information Service (Seoul TOPIS) covers four days for two weekdays (13th and 14th December 2013) and two weekends (16th and 17th December 2013) and contains the speed, angle, and position identified by the global positioning system (GPS) for an individual taxi; the temporal resolution is 10 seconds, and the spatial resolution is about 5~\si{m} to 10~\si{m}.

\begin{figure}
\includegraphics[width=1\linewidth]{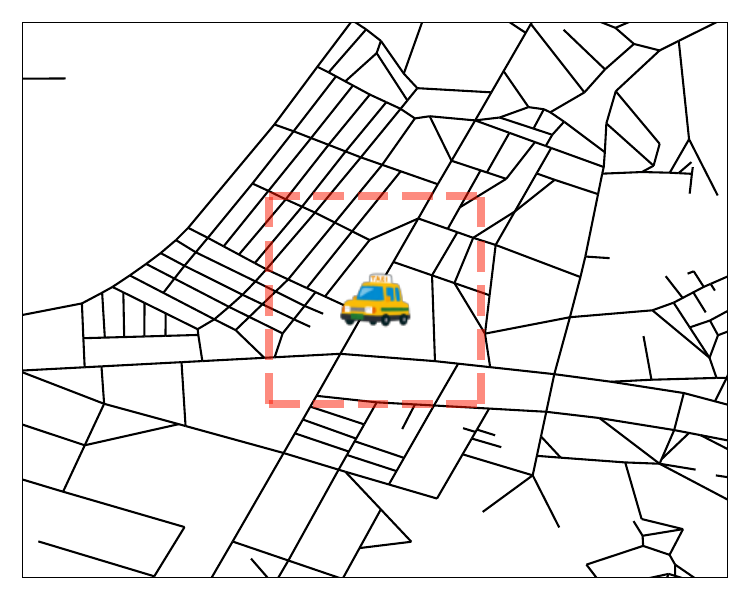}
\caption{The speed matching process into adjacent roads. When the TOPIS data provides a speed value $v_i$ about the taxi on the road $i$ symbolized by the icon, create a square cell (represented by the red dashed square) around the focal taxi. Let us call such a road $i$ a source road. Assign the speed $v_i$ into every adjacent road $j$ in a cell, if a speed value of the road $j$ is not contained in the TOPIS. If the road $j$ belongs to multiple cells by several source roads, then the average speed over the speeds of the source roads is assigned as $v_j$. We set the linear length of the box to be 300~\si{m}.
}\label{fig:taxi}
\end{figure}

Several good algorithms for mapping such GPS information of vehicles onto road networks~\cite{li2013large, yang2018fast, guo2019urban} are not suitable to our case due to a scarcity of the TOPIS data~\cite{note}, so we furnish a simple method as follows. Create a square cell of 300\si{m}$\times$300\si{m} centred around an individual taxi.  Then, assign a speed $v$ of a taxi to all the links (roads) inside the cell (see Fig.~\ref{fig:taxi}). When some links are covered by the adjacent multiple cells (formed by the multiple taxis), the links have the average speeds of the taxis as the link weights. For the sake of convenience, we take the lower temporal resolution of 1 hour, then the temporal average of the speed within a given time interval is also taken into account. In addition, we aggregate the speed on either weekdays (13th and 14th) or weekends (16th and 17th) to smooth fluctuation out for hour analysis. Hence, the speed $v_i(t; d)$ for road $i$ for a given day $d$ (the ``day'' herein corresponds to either the two-weekday or two-weekend) is the averaged value over the GPS speeds adjacent to the road $i$, the time window with 1 hour, and two-day information (weekday or weekend). After completing such a matching process, the basal network for the percolation analysis is reduced to 42,874 nodes and 59,965 links for weekdays and to 42,226 nodes and 58,702 links for weekends. 
 

\subsection{Rescalde speed for traffic-status evaluation}
\begin{figure*}
\includegraphics[width=1\linewidth]{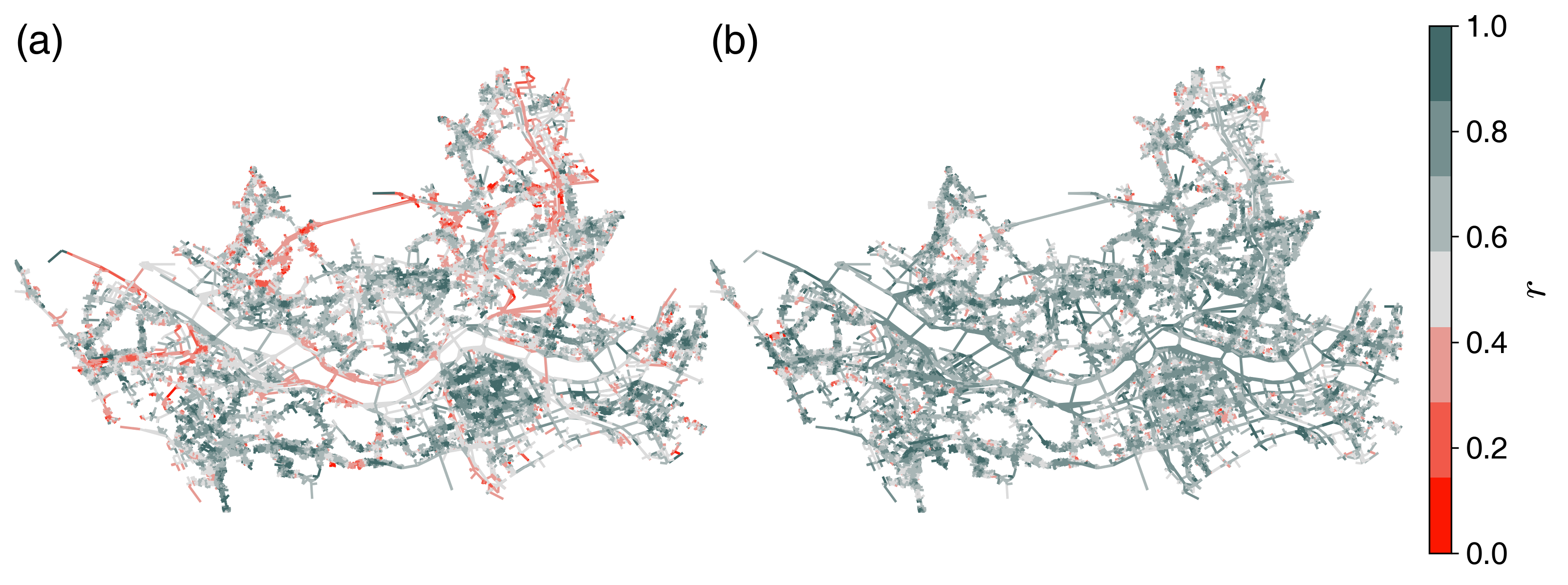}
\caption{The rescaled speed $r$ distributed the basal road network of Seoul with 42,874 nodes (intersections) at (a) 8:00 (morning rush hour) and (b) 22:00 (non-rush hour) on two-weekday (13th and 14th Dec. 2013). The closer to 0 (1) the value of $r$ is, the more congested (smoother) the road is, respectively. 
}\label{fig:r_on_roads}
\end{figure*}

There are various types of urban roads, e.g., local streets, alleyways, and highways. They have their own characteristics and roles, causing the different traffic capacities of the roads based on the speed limit and the number of speed bumps per length. Given these features, the relative speeds of vehicles rather than the absolute speeds can better stand for assessing the traffic status such as being smooth or congested. Instead of the speed $v_i(t; d)$ on a road $i$ assigned in Sec.~\ref{subsec:road_taxi}, we exploit a \emph{rescaled speed} $r_i(t; d)$ at time $t$ for a given day $d$ to describe the traffic status~\cite{li2015percolation}, computed as 
\begin{equation}
    r_i(t; d) \equiv \frac{v_i (t; d)}{v_i^{\rm max}(d)},
\label{eq:r}
\end{equation}
and $v_i^{\rm max}(d)$ was defined as the top 5 \% speed on road $i$ for a given day $d$ in the previous study~\cite{li2015percolation}. We refine this approach by defining $v_i^{\rm max}(d)$ as the second-fastest speed, with the 5\% threshold positioned between the first and second ranks among the twenty-four entities. To blur out the effect of some outliers, we adopt the effective maximum $v_i^{\rm max}(d)$ rather than the real maximum speed, e.g. $\max[v_i(t; d)]$, and assign the effective maximum speed to $v_i (t;d)> v_i^{\rm max}(d)$. It results in the rescaled range of $r$ up to unity. Figure~\ref{fig:r_on_roads} showcases the rescaled speed $r$ across the basal road network obtained from Sec.~\ref{subsec:road_taxi}, at a morning rush hour (8:00) and at a non-rush hour (22:00) on the two-weekday (average of 13th and 14th Dec. 2013). At 8:00, the peripheral parts of the roads start to be congested (low $r$); many commuters dwelling in the neighboring metropolitan area go to their workplace mostly located in the central area of Seoul [Fig.~\ref{fig:r_on_roads}(a)]. Otherwise, during the off-peak hour (22:00), the traffic status looks quite smooth with little movement along the roads [Fig.~\ref{fig:r_on_roads}(b)]. One can also sense that some congested or smooth traffic clusters form across the road networks. The distribution of the clusters allows us to assess the overall traffic situation such as a dominance of congestion or free-flow, so we take a percolation approach for the cluster analysis.

\section{Traffic clusters as percolation}
\label{sec:percolation}

We adopt the percolation analysis to identify the robustness of the free-flow cluster~\cite{li2015percolation, zeng2019switch}. Introducing the threshold value $q$, the congested links (road) having smaller $r$ than $q$ are removed from the network. As $q$ increases, the free-flow clusters shrink more by ruling out the least smooth roads. The collapse of the free-flow clusters is not the same as the formation of the congestion clusters although they are definitely related; they are not the exclusive process, so they should be separately treated if needed~\cite{kwon2024clusterformationsfreecongested}. 

\subsection{Breakdown of the giant connected component in the free-flow cluster}
\label{subsec:gc}

We analyze the behavior of the GCC of the free-flow cluster, equivalently describing the prevalence of the free flows, against removing a link with $r<q$. The threshold $q$ increases by $\Delta q=1/1000$ from $q=0$ and the number of links to be removed between $q \leq r < q+\Delta q$ can differ depending on the value of $q$ due to the non-uniform distribution of $r$ (the distributions have very narrow forms in our case, although they are not shown in this paper). The percolation method in traffic networks follows the method proposed in Refs.~\cite{li2015percolation, zeng2019switch} (see the consideration of the adaptive $\Delta q$ depending on the number of links between $q$ and $q+\Delta q$
~\cite{kwon2024clusterformationsfreecongested}). The previous studies have not explicitly addressed correlation effects emerging on networks, which we aim to explore together in this work. 

We measure the relative size $s_G$ of the giant connected component that can portray the extent of the prevalence of congestion over the roads. It is computed as
\begin{equation}
    s_\mathrm{G}(q) = \frac{S_1(q)}{N},
\label{eq:sg}
\end{equation}
where $N=42,874$ for weekdays or $N=42,226$ for weekends is the total nodes of the basal network, $S_1(q)$ is the number of the nodes (intersections) belonging to the GCC at a given threshold $q$.

\begin{figure*}
\centerline{\includegraphics[width=0.75\linewidth]{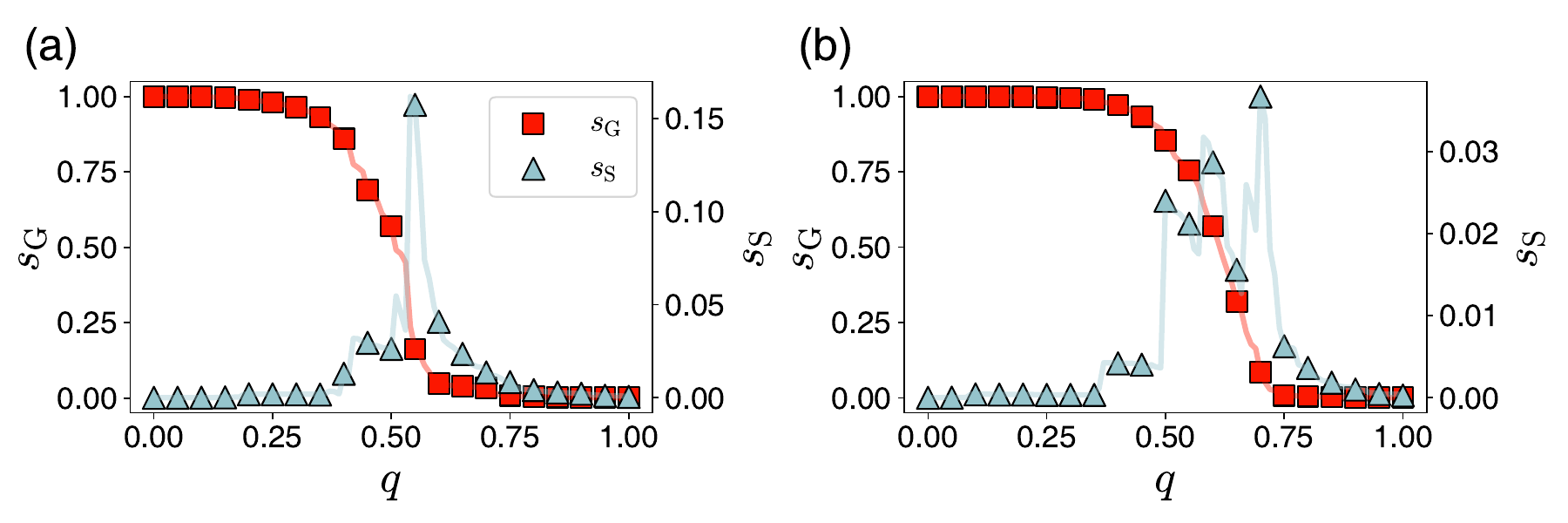}}
\caption{
The robustness of the connected components consists of free-flow roads (a) at 8:00 (morning rush hour) and (b) at 22:00 (non-rush hour) on weekdays. The relative size $s_\mathrm{G}$ [Eq.~(\ref{eq:sg})] of the GCC and the $s_\mathrm{S}$ [Eq.~(\ref{eq:ss})] of the SCC are measured against the link removal with the rescaled speed $r$ less than the threshold $q$. The $s_\mathrm{S}$ peaks when the $s_\mathrm{G}$ decreases sharply, and we define such $q$ as the transition point $q_c$.
}
\label{fig:gccscc}
\end{figure*}

As representative cases, Fig.~\ref{fig:gccscc} displays the behaviors of the GCC during the morning rush hour (8:00) and non-rush hour (22:00) on weekdays. The free-flow GCC endures against the removal and then collapses into smaller components as the threshold $q$ increases. The decreasing behavior (e.g. slope) of the GCC shows an insignificant difference between the morning rush hour and non-rush hour, but the certain point where the remarkable breakdown of the GCC occurs is more delayed at 22:00 than at 8:00. We call such the certain point the transition (threshold) point $q_c$, which can stand for the endurance of the free-flow GCC against the breakdown~\cite{zeng2019switch}. 

Aiming to both investigate the formation of the connected clusters and identify the $q_c$, we examine the second connected component (SCC) as
\begin{equation}
    s_\mathrm{S}(q) = \frac{S_2(q)}{N},
\label{eq:ss}
\end{equation}
where the $S_2$ is the number of nodes belonging to the SCC. The SCC curve during the morning rush hour has only one dominant peak, whereas multiple peaks are observed during the non-rush hour. While the GCC breaks down, many smooth roads and intersections are still concentrated not forming any significant small clusters during the rush hour, which can be deemed as the easy collapse of the free-flow GCC as ever [Fig.~\ref{fig:gccscc}(a)]. On the other hand, the secondary dominant clusters start to emerge during the non-rush hour  [Fig.~\ref{fig:gccscc}(b)].  Even though the breakdown behavior of the GCC in Eq.~(\ref{eq:sg}) within some interval of $q$ is similar between rush and non-rush hours, its effective dominance of the giant free-flow seems quite different, characterized by the SCC's emergence pattern.

\begin{figure}[b]
\includegraphics[width=1\linewidth]{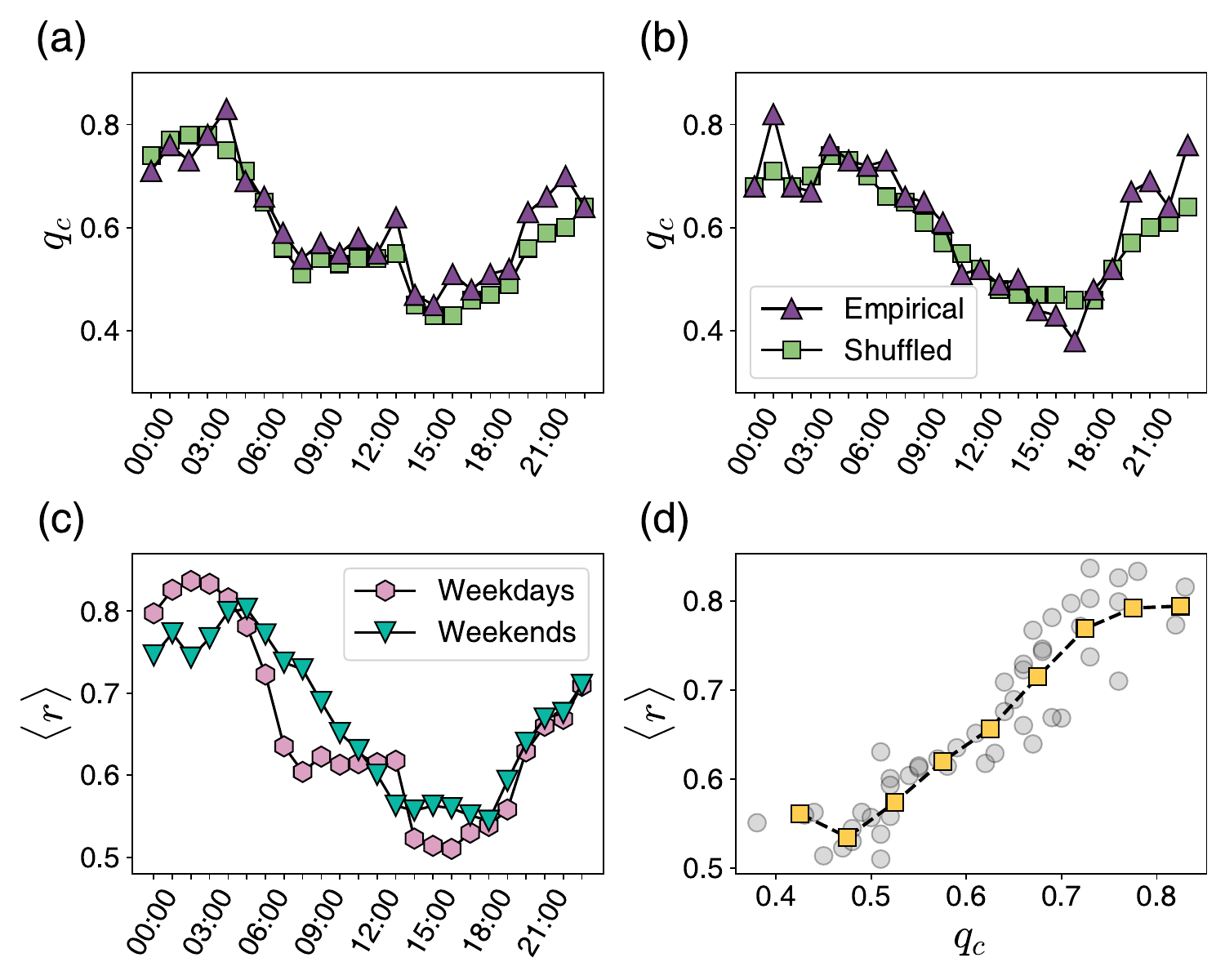}
\caption{Temporal behaviors of the transition point $q_c$ and relation with the average rescaled speed $\langle r \rangle$. The $q_c$'s obtained from the empirical data (triangle) and weight-shuffled networks (square) are displayed every hour (a) on weekdays and (b) on weekends. (c) The mean rescaled speed $\langle r \rangle$'s on weekdays (hexagon) and weekends (inverted triangle) are plotted every hour. (d) The $\langle r \rangle$ versus $q_c$ for the empirical data both on weekdays and weekends are in the scatter plot altogether, symbolized by the gray circles, with a Pearson correlation coefficient of $0.9123$. The yellow squares represent the average values.
}
\label{fig:qc_meanr}
\end{figure}

It could be imprecise to identify the transition point in empirical data due to its inherent noise, so we exploit the quantification method based on percolation theory; the point $q_c$ is practically defined as the point at which the SCC in Eq.~(\ref{eq:ss}) has the maximum, i.e., 
\begin{equation}
q_c\equiv \arg\max_{q} s_\mathrm{S}(q).    
\label{eq:qc}
\end{equation}
The temporal behaviors of $q_c$ for weekdays and weekends are exhibited in Figs.~\ref{fig:qc_meanr}(a) and~\ref{fig:qc_meanr}(b). The large $q_c$ indicates the long endurance and high robustness of the free-flow GCC and vice versa. 

\begin{figure*}
\includegraphics[width=\linewidth]{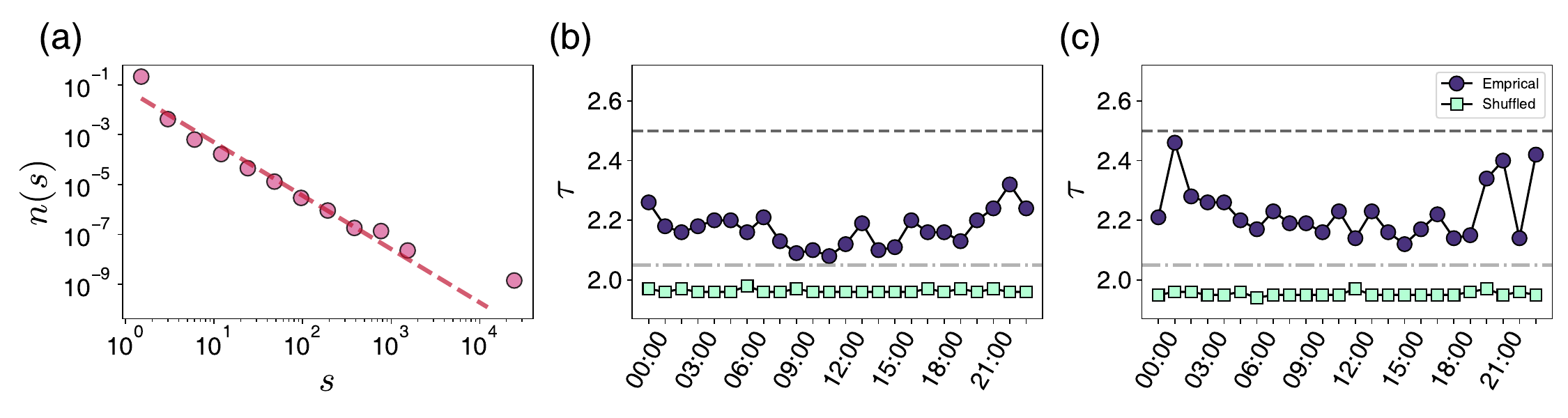}
\caption{The distribution $n(s)$ of the cluster size $s$ and its critical exponent $\tau$. (a) At 8:00 (morning rush hour) on weekdays, the exponent value is estimated as $\tau\simeq 2.14$. The temporal behaviors of $\tau$ for (b) weekdays and (c) weekends are displayed. The circle and the square symbols represent $\tau$'s obtained from the empirical data and the shuffled networks, respectively. The two dashed lines indicate $\tau^{\rm (MF)}=5/2$ for the mean field and $\tau^{\rm (2D)}=187/91$ for two-dimensional space, belonging to the universality classes in the ordinary percolation.}
\label{fig:cluster}
\end{figure*}

On weekdays in Fig.~\ref{fig:qc_meanr}(a), $q_c$ in the morning remains high with a decreasing pattern and reboundes during the evening. The $q_c$ peaks at 4:00 (early morning). People begin their daily work and utilize the transportation infrastructure, including vehicles, from this early morning, leading to the free-flow cluster's breakdown. In parallel with this human activity, the transition point $q_c$ also decreases, indicating less robustness of the free-flow clusters. This declining trend is maintained until the morning rush hour. There is a slight increase of $q_c$ after the morning congestion's release, and the increase is repeated after evening rush hours following the decrease of $q_c$ from lunchtime to the rush hours. The timeslots with decreasing and increasing $q_c$ (thus, the less and more robustness of the free-flow clusters against link removal) are nearly consistent with the typical severe and relaxed congestion, respectively.

The overall behavior, except for the slight increase during lunchtime, on weekdays is similarly observed on weekends [Fig.~\ref{fig:qc_meanr}(b)], but a comparable level of $q_c$ appears at a slightly later time, including the lower level at lunchtime differently from the weekdays; it could be equivalent to the later beginning of the human's daily life on weekends than weekdays, including that people move to start from lunchtime. Embracing similar and dissimilar points simultaneously, human activity can affect the usage pattern of the road, resulting in the traffic cluster's breakdown characterized by $q_c$. 

To identify the correlation effect on determining the transition point $q_c$, we compare it to uncorrelated cases. In doing so, we shuffle all weights [i.e., rescaled speed $r$ in Eq.~(\ref{eq:r})] over the basal networks to generate 100 uncorrelated networks and obtain the averaged $q_c$ over the 100 networks. As displayed in Figs.~\ref{fig:qc_meanr}(a) and~\ref{fig:qc_meanr}(b), the empirical $q_c$ and the shuffled (average) $q_c$ has insignificant difference. Due to the unimodal distribution of $r$, this percolation method exhibits low removal resolution near the critical threshold where the GCC undergoes significant fragmentation. Consequently, the distinction between the empirical and uncorrelated cases becomes indistinguishable. However, if one considers other types of percolation, e.g., $q$ being the link fraction of occupation/removal~\cite{kwon2024clusterformationsfreecongested}, the transition threshold $q_c$ serves as a key indicator for differentiating traffic behavior based on correlation effects.

To back the role of $q_c$, we further investigate the mean relative speed as
\begin{equation}
    \langle r \rangle (t; d) \equiv  \frac{1}{L}\sum_{i} r_i (t; d),
\label{eq:meanr}
\end{equation}
where $L$ is the number of links (roads) as seen in Fig.~\ref{fig:qc_meanr}(c). The whole shapes of the curves of $\langle r \rangle$ are almost identical to those of the $q_c$'s each. The close relationship between $\langle r \rangle$ and $q_c$ is fairly well captured in the scatter plot in Fig.~\ref{fig:qc_meanr}(d), with a high level of the Pearson correlation coefficient (about 0.9). From these observations, it seems quite persuasive that the high $q_c$ (long endurance of the free-flow GCC) symbolizes the smooth traffic situation signalled by the high $\langle r \rangle$.


\subsection{Distribution of the free-flow clusters}
\label{subsec:cluster}

The previous study~\cite{zeng2019switch} has uncovered that, for three metropolitan cities in China, the distribution $n(s)$ of the free-flow clusters follows the power law at the transition point $q_c$:
\begin{equation}
    n(s; q_c) \sim s^{-\tau}, 
\end{equation}
where $s$ is the cluster size and $\tau$ is the Fisher exponent. The exponent $\tau$ is measured using the linear regression on the log-binned distribution. Their finding also indicates that the exponent value is bounded as $\tau^{\rm (2D)} \lesssim \tau \lesssim \tau^{\rm (MF)}$ and contingent upon the road structures and traffic situation.  If outer ring roads that serve as shortcuts are actively used, $\tau$ is closer to $\tau^{\rm (MF)}=5/2$, belonging to the mean-field universality class; otherwise, $\tau$ approaches $\tau^{\rm (2D)}=187/91$, corresponding to the two-dimensional lattice. We first observe the similar scaling behavior for Seoul during the morning rush hour (8:00) on weekdays as seen in Fig.~\ref{fig:cluster}(a). The evaluated exponent is $\tau\simeq2.14$, which means that the shortcuts are hard to perform a useful function as a result of the congestion, during the rush hour. To complement the observation in terms of correlation effect, we will compare the shuffled cases generated in Sec.~\ref{subsec:gc}. 


We explore the temporal patterns of the exponent $\tau$ on weekdays [Fig.~\ref{fig:cluster}(b)] and on weekends [Fig.~\ref{fig:cluster}(c)]. Albeit the temporal fluctuations of $\tau$ appear less pronounced compared with the transition point $q_c$ in Figs.~\ref{fig:qc_meanr}(a) and~\ref{fig:qc_meanr}(b), the basic trend persist: $\tau$ is lower during periods of active human movement (morning to evening) and higher otherwise (midnight to daybreak). In other words, when the free-flow GCC is robust (induced by low human activity) involving a hard emergence of the SCC, the traffic network behaves as if it were embedded in the effectively high-dimensional space, efficiently using shortcuts like highways. When congestion is dominant on the roads (induced by high human activity), the free-flow clusters are prone to segmentation into small clusters, resulting in the behavior in the effectively low-dimensional space. However, the distinction between rush and non-rush hours from morning to evening remains unclear due to the low temporal variability of $\tau$.

\begin{figure*}[t]
\includegraphics[width=0.75\linewidth]{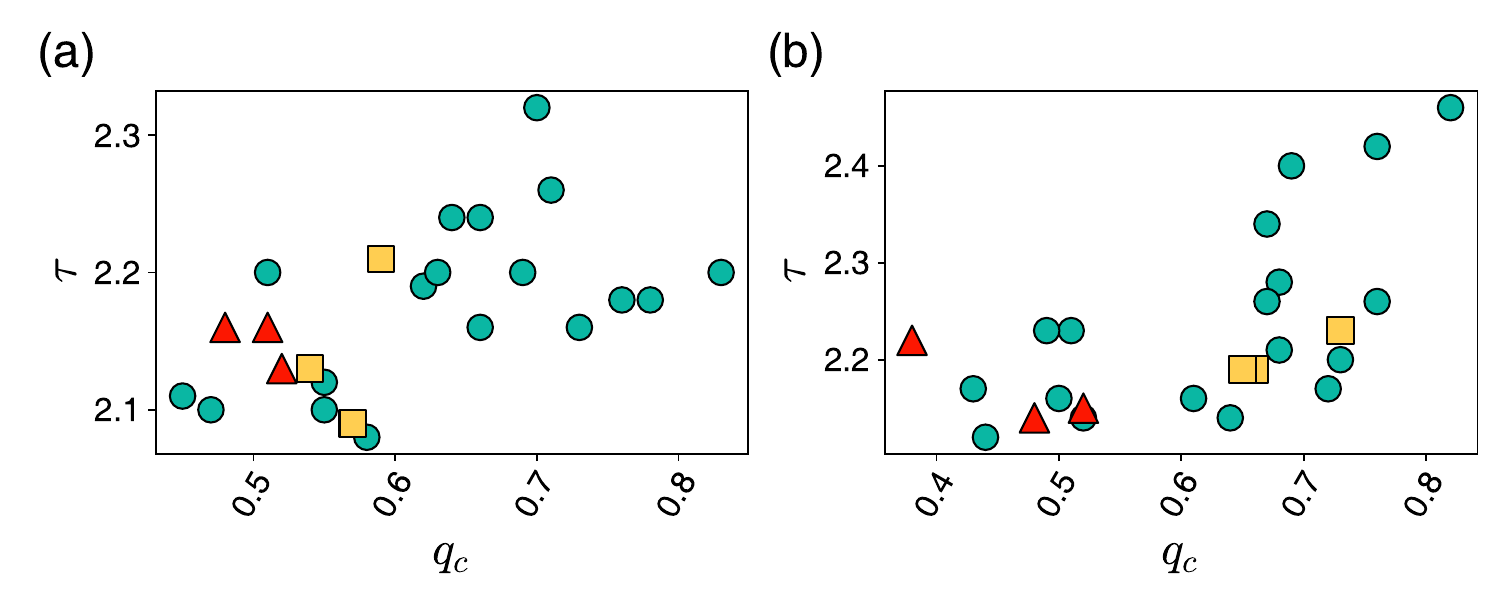}
\caption{Scatter plots of the transition point $q_c$ and the Fisher exponent $\tau$ are shown for (a) weekdays and (b) weekends. Red triangles and yellow squares represent typical commuting times in the morning (7:00, 8:00, and 9:00) and evening (17:00, 18:00, and 19:00), respectively. While commuting hours correspond to rush hours on weekdays, they do not necessarily indicate rush periods on weekends. (a) On weekdays, both $q_c$ and $\tau$ tend to be relatively low during rush hours, indicating an earlier breakdown and easier fragmentation of the free-flow GCC. Conversely, both metrics are generally higher during non-rush hours.  (b) On weekends, a similar trend is observed, where high $(q_c, \tau)$ values correspond to periods of lower human activity, and vice versa.}
\label{fig:qc_tau}
\end{figure*}

The dispersion of $\tau$ appears more mitigated compared to $q_c$, suggesting reduced variability. The association between $q_c$ and $\tau$ provides a more integrated understanding of the traffic system in Seoul, offering deeper insights than when these metrics are analyzed separately. Figure~\ref{fig:qc_tau} presents the correlation, revealing a positive relationship. Specifically, both $q_c$ and $\tau$ are lower during rush hours and higher during non-rush hours, as shown in Fig.~\ref{fig:qc_tau}(a). During the afternoon non-rush period, some $(q_c,\tau)$ pairs exhibit lower values, likely due to the limited sample size. We anticipate that with a larger dataset, the distinction between rush and non-rush hours would become more pronounced, consistent with the findings of a previous study~\cite{zeng2019switch}. Note that there is no typical commuting time on weekends [Fig.~\ref{fig:qc_tau}(b)].

For the uncorrelated cases, $\tau$ remains nearly constant, making a significant difference from the empirical data, unlike the behaviors of $q_c$. Its average and standard error are 1.96 and 0.006 for weekdays and 1.95 and 0.007 for weekends, respectively, and the averages are remarkably less than $\tau^{\rm (2D)}=187/91$ and rather close to the Fisher exponent in the one-dimensional space (i.e., $\tau^{\rm (1D)}=2$). For the shuffled cases, it is difficult to clearly understand $\tau<\tau^{\rm (1D)}$, but one may first suspect that the following factors are responsible for the low value less than $\tau^{\rm(2D)}$: the effective dimension of the basal network (from the structural point of view) and the distribution of the weight $r$ (from the functional point of view).

The basal road networks are planar networks embedded in nearly two-dimensional space on the earth's surface, but their geometry is inherently less regular. These networks often form long loops, reducing their effective embedding dimension to less than 2. Moreover, the percolation process involving deterministic link removal differs from the ordinary percolation method, particularly due to its dependence on the distribution pattern of $r$. The distribution of $r$ is unimodal, signaling the bulk elimination of $r$ around the mean value $\langle r\rangle$ according to our percolation method.
When a large number of very similar $r$'s are randomly scattered on the network without clustering unlike the empirical data, their simultaneous elimination can significantly interfere with the connected structures, further reducing the effective dimension functionally. These structural and functional factors could reduce the effective dimension, resulting in $\tau<\tau^{\rm (2D)}$ for the uncorrelated case. However, the Fisher exponent less than $\tau^{\rm (1D)}=2$ is still not fully understood, which may be caused by the finite-size effect. 

\section{CONCLUSIONS}
\label{sec:conclusion}

In this study, we have explored the traffic percolation of Seoul as a case study, in parallel with previous research for three metropolitan cities in China~\cite{zeng2019switch}. We have obtained the results that the transition point $q_c$ [Eq.~(\ref{eq:qc})] and the Fisher exponent $\tau$ characterize well the road usage patterns based on human's temporal and daily activity, in line with the previous study. In addition to these findings, we have measured the mean rescaled speed $\langle r \rangle$ for consolidating the meaning of $q_c$ and introduced the shuffled cases for analyzing correlation effects. We have also found that the weight-weight correlation does not affect the transition point $q_c$ but makes a significant difference in the Fisher exponent $\tau$, and suggested that $\tau\simeq 1.96$ obtained from the shuffled cases become the lower baseline rather than $\tau^{\rm (2D)}=187/91$. We have guessed that the baseline $\tau<\tau^{\rm (2D)}$ arises from both the structural properties and the functional characteristics of the traffic network, particularly the distribution of $r$. Additionally, the finite-size effect may further contribute to the baseline $\tau$, making it even lower than $\tau^{\rm (1D)}$. When focusing on the empirical case, the combined analysis of $q_c$ and $\tau$ provides a more comprehensive understanding of the traffic system than considering these metrics individually. 

This study has several limitations. To ensure a consistent comparison of urban characteristics under the same traffic analysis framework, we adopted the percolation method used in a previous study~\cite{zeng2019switch}. However, the estimation of key metrics such as $q_c$, $s_{\rm G}$, and $\tau$ is sensitive to the distribution of $r$. Since the percolation method employed here removes a varying number of links for a given $q$, the functional form of the $r$ distribution may influence the qualitative results. To mitigate this dependency, an alternative method based solely on the rank of $r$ has been proposed in a separate study~\cite{kwon2024clusterformationsfreecongested}, which has distinguished $q_c$ between the empirical and uncorrelated cases, in contrast to the approach used here. In addition, our analysis of the Fisher exponent was compared only to ordinary percolation models, while real-world traffic networks exhibit inherent weight-weight correlations. It would therefore be valuable to reconsider these results in the context of correlated percolation~\cite{SABERI20151, prakash1992structural, chalhoub2024, zeng2019switch}. Finally, this study focuses exclusively on Seoul, and future research could explore whether similar patterns emerge across metropolitan cities in different countries, provided that sufficient data is available.

The comparison with the shuffled cases in this study offers a practical perspective on understanding and managing urban traffic networks. By providing a baseline comparison, the shuffled cases allow us to isolate the influence of real-world correlations, such as spatial and temporal dependencies, on traffic dynamics. This approach highlights how natural clustering in traffic patterns, often driven by geographic constraints or regulatory practices, impacts congestion and the robustness of road networks. Furthermore, the shuffled cases serve as a benchmark for evaluating intervention strategies, enabling the identification of critical roads or intersections that require reinforcement or optimization. They also enhance predictive modeling by refining our ability to simulate correlated and uncorrelated traffic scenarios, offering valuable insights for forecasting the effects of systemic changes like road closures or infrastructure upgrades. Lastly, shuffled cases reveal the baseline resilience of networks to random disruptions, guiding the design of more robust and efficient urban traffic systems. These findings underscore the value of incorporating shuffled cases as a practical tool in traffic planning, policy-making, and sustainable urban development.

\begin{acknowledgments}
This work was supported by the National Research Foundation (NRF) of Korea through Grant Numbers. NRF-2023R1A2C1007523 (S.-W.S.) and RS-2024-00341317 (M.J.L.). This work was also partly supported by the Institute of Information \& communications Technology Planning \& Evaluation (IITP) grant funded by the Korean government (MSIT) (No.RS-2022-00155885, Artificial Intelligence Convergence Innovation Human Resources Development (Hanyang University ERICA)). We sincerely thank the Seoul Transport Operation and Information Service (Seoul TOPIS) for providing the data for this study.
\end{acknowledgments}


\begin{thebibliography}{34}
\bibitem[{not()}]{note}
\bibinfo{note}{In the TOPIS data, taxis cover only a small portion of the entire road network. GPS data may be incomplete due to limitations in temporal or spatial resolution. As a result, nonzero weights (relative speeds) are assigned to only a fraction of the links (roads) in the network, potentially underrepresenting the underlying structure of the network on which the percolation process is applied.}
\end{thebibliography}

\end{document}